\documentclass[universe,article,accept,pdftex,moreauthors]{Definitions/mdpi} 
\preto{\abstractkeywords}

\firstpage{1} 
\pubvolume{11}
\issuenum{11}
\articlenumber{86}
\pubyear{2025}
\copyrightyear{2025}
\externaleditor{Laszlo Szabados and Tamas Szalai} 
\datereceived{6 February 2025} 
\daterevised{3 March 2025} 
\dateaccepted{5 March 2025} 
\datepublished{6 March 2025} 
\hreflink{https://doi.org/10.3390/\\ universe11030086} 

\usepackage{url}
\usepackage{supertabular}
\usepackage{geometry}
\usepackage{amsmath}
\usepackage{longtable}
\usepackage{supertabular}

\usepackage{amsfonts}
\usepackage{amsmath}
\usepackage{mathrsfs}
\usepackage{epsfig}

\def\aap{Astron. Astrophys.}

\def\apj{Astrophys. J.}
\def\apjl{Astrophys. J. Lett.}

\def\aj{Astron. J.}
\def\mnras{Mon. Not. R. Astron. Soc.}

\def\apss{Astrophys. Space Sci.}

\def\aaps{Astron. Astrophys. Suppl.}
\def\pasp{Publ. Astron. Soc. Pac.}

\def\raa{Res. Astron. Astrophys.}
\def\jrasc{J. R. Astron. Soc. Can.}
\def\jaavso{ J. Am. Assoc. Var. Star Obs.}
\def\nar{New Astron. Rev.}
\def\memsai{Mem. Soc. Astron. Ital.}

\Title{Period Variation Rates of Four Radial Single-Mode High-Amplitude Delta Scuti Stars}

\TitleCitation{Period Variation Rates of Four Radial Single-Mode High-Amplitude Delta Scuti Stars}

\Author{{Tian-Fang Ma} $^{1}$, {Jia-Shu Niu}  $^{1,2,3,}$*\orcidA{} 
and {Hui-Fang Xue} $^{4,5,6}$\orcidB{}} 

\AuthorNames{Tian-Fang Ma, Jia-Shu Niu, and Hui-Fang Xue }

\isAPAStyle{%
       \AuthorCitation{Ma, T.-F., Niu, J.-S., \& Xue, H.-F.}
         }{%
        \isChicagoStyle{%
        \AuthorCitation{Tian-Fang Ma, Jia-Shu Niu, and Hui-Fang Xue.}
        }{
        \AuthorCitation{Ma, T.-F.; Niu, J.-S.; Xue, H.-F.}
        }
}

\address{%
$^{1}$ \quad Institute of Theoretical Physics, Shanxi University, Taiyuan 030006, China; \\ 
$^{2}$ \quad State Key Laboratory of Quantum Optics Technologies and Devices, Shanxi University, \linebreak Taiyuan 030006, China\\
$^{3}$ \quad Collaborative Innovation Center of Extreme Optics, Shanxi University, Taiyuan 030006, China\\
$^{4}$ \quad Department of Physics, Taiyuan Normal University, Jinzhong 030619, China; \\
$^{5}$ \quad Institute of Computational and Applied Physics, Taiyuan Normal University, Jinzhong 030619, China\\
$^{6}$ \quad Shanxi Key Laboratory for Intelligent Optimization Computing and Blockchain Technology, \linebreak Jinzhong 030619, China\\
}

\corres{Correspondence: jsniu@sxu.edu.cn}

\abstract{In this work, we present a study on the long time-scale period variations of {four} 
single-mode high-amplitude delta Scuti stars (HADS) via the { classical {$O-C$}} 
analysis. The target { HADS} are (i) XX Cygni, (ii) YZ Bootis, (iii) GP Andromedae, and (iv) ZZ {Microscopii}. { The newly determined times of maximum light came from the Transiting Exoplanet Survey Satellite (TESS), American Association of Variable Star Observers (AAVSO), and Bundesdeutsche Arbeitsgemeinschaft f$\ddot{\rm u}$r Ver$\ddot{\rm a}$nderliche Sterne (BAV) projects. Together with the times of maximum light obtained in the historical literature, the $O-C$ analysis was performed on these HADS, in which we obtained the linear period variation rates $\dot{P}/P$ as $(9.2 \pm 0.2) \times 10^{-9} \ \mathrm{yr^{-1}}$,  $(3.2\pm 0.2)\times 10^{-9} \ \mathrm{yr^{-1}}$, $(4.22\pm 0.03) \times 10^{-8} \ \mathrm{yr^{-1}}$, and $(-2.06 \pm 0.02) \times 10^{-8} \ \mathrm{yr^{-1}}$, respectively.} 
Based on these results and some earlier research, we also discuss the {evolutionary stages} and the mechanisms of the period variation of these four HADS.
}

\keyword{pulsating stars; delta Scuti stars; stellar evolution}

\begin{document}

\section{Introduction}

Delta Scuti stars are pulsators located in the classical Cepheid instability strip {crossing the main sequence and sometimes in the region between the main sequence and the giant branch}. The mass of delta Scuti stars {ranges} from 1.4 $M_{\odot}$ to 2.5 $M_{\odot}$, and their pulsation period {ranges} from $0.02$ to $0.3$ days \cite{2022MNRAS.515.4574N}. 
{As a subclass of delta Scuti stars, the High Amplitude Delta Scuti stars (HADS) generally have larger amplitude ($\ge$0.1 mag) and slow rotation speed ($v\sin{i} \leq 30\ \mathrm{km\ s^{-1}}$) \cite{2000ASPC..210....3B}, which can be Population I or II stars.}
Most of the HADS exhibit one or two radial pulsation modes, while some of them display three radial pulsation modes {and even} some non-radial modes \cite{2008A&A...478..865W,2013RAA....13.1181N,2017MNRAS.467.3122N,2018ApJ...861...96X,2020ApJ...904....5X,2022MNRAS.512.3551D}.  For those HADS which exhibit only one{ dominant radial pulsation mode (see, e.g., \cite{2023RAA....23g5002X}), we can use the deviation of their times of maximum light (TML) between the observed and calculated ones to study their period variations ($O-C$ analysis).}

{The $O-C$ analysis provides us an opportunity to study the period variation details of these stars. Theoretically, for HADS, the period variation rates caused by stellar evolution should be in the range of $10^{-10}\ \mathrm{yr^{-1}} $ to $10^{-7}\ \mathrm{yr^{-1}} $ \cite{1998A&A...332..958B}.} If the observed value is of the same order of magnitude as the theoretical one, it {indicates} that the period variation of this star could be attributed to stellar evolution; if not, the period variation may be influenced by {other} mechanisms such as nonlinear mode interaction, mass transfer or the {light-travel time effect in binary systems} \cite{1998A&A...332..958B,2016MNRAS.460.1970B,2020ApJ...904....5X,2021MNRAS.504.4039B,2017MNRAS.467.3122N,2018ApJ...861...96X,Niu_2022,Xue_2022}. Consequently, the investigation into the period variation of HADS will help us have a better understanding of their {evolutionary stage} and reveal {the} unknown mechanisms behind it.

In this work, we chose {four single-mode HADS as the targets to study their period variation rates, which have been observed by different projects many times in history and recent years.}   
The basic information about these four HADS {is listed} as follows.

\textls[-25]{{XX Cygni} (XX Cyg, $\langle V \rangle = 11^{\rm m}.7$, {$P_0 \approx 0.1349\ {\rm d}$}) \cite{2012AJ....144...92Y}, a Population II HADS \cite{1904AN....165...61C,Joner_1982,Hintz_1997}. The 
linear period variation rate $\dot{P}/P$ {obtained by refs. \cite{2012AJ....144...92Y,2021JRASC.115..238W}} are both at the order $10^{-8}\ \mathrm{yr}^{-1}$.}

{YZ Bootis} (YZ Boo, $\langle V \rangle = 10^{\rm m}.57$, {$P_0 \approx 0.1041\ {\rm d}$}) \cite{2018RAA....18....2Y}, a Population I HADS \cite{1983PASP...95..433J}. In the {most recent $O-C$ analysis on YZ Boo} \cite{2018RAA....18....2Y}, the authors {obtained} a period variation rate of $\dot{P}/P$ $=$ $ 6.7(9) \times 10^{-9}\ \mathrm{yr}^{-1}$. However, they {also declared} that the second-order {polynomial} curve fitting does not show {sufficient advantage compared to} linear fitting, which {indicates that it is hard to say whether the period of YZ Boo is varying or not.} 

\textls[-10]{{GP Andromedae} (GP And, $\langle V \rangle = 10^{\rm m}.7$, {$P_0 \approx 0.0787\ {\rm d}$}) \cite{2011AJ....142..100Z}, a Population I HADS \cite{1998A&A...332..958B}.
In the latest research of {its} period variation rate, {ref.} \cite{2011AJ....142..100Z} {not only gave a result of $\dot{P}/P$ = $ (5.49\pm 0.1)\times 10^{-8} \ \mathrm{yr^{-1}}$, but also confirmed an amplitude variability in GP And.}
{Two other} recent studies for GP And \cite{2006IBVS.5718....1S,pop2003period} showed increasing pulsation period variations with a rate of $\dot{P}/P$ $\sim$ $ 6 \times 10^{-8}\ \mathrm{yr}^{-1}$. All these results are much {larger} than the theoretical values \cite{rodriguez1993simultaneous}.}

{ZZ Microscopii} (ZZ Mic, $\langle V \rangle = 9^{\rm m}.43$, {$P_0 \approx 0.0672\ {\rm d}$}), was first {discovered} by \cite{1961Obs....81...25C}. {In ref.} \cite{2009PASP..121..478K}, they conducted an $O-C$ analysis on ZZ Mic and obtained that its period {was increasing at} a constant rate. However, in the {latest} $O-C$ research of this star \cite{2015JAVSO..43...50A}, the authors claimed that the period of ZZ Mic was  increasing at a constant rate during the years 1960 to 2003, and it {has been decreasing} since 2003.

All of these stars have unsolved problems about their period variations. Thus, we {expected to  acquire a better understanding on the period variations of them with more data.} This paper is {organized} as follows. In Section \ref{sec:fr}, we introduce the data sources and the reduction procedures. In Section \ref{sec:O-C}, we show the $O-C$ results of four target stars and their {linear} period variation rates.
In Section \ref{sec:discussion}, we {discuss} the $O-C$ behaviors and {the origins of the period variation rates}. In Section \ref{sec:conclusions}, we {list the conclusions}.

\section{Data Sources and Data Reduction}
\label{sec:fr}
{The Transiting Exoplanet Survey Satellite (TESS), a NASA mission launched in 2018, is designed to discover exoplanets by detecting transits while simultaneously advancing asteroseismology through its high-precision photometric observations of host stars, enabling scientists to study stellar interiors and oscillations for deeper insights into stellar evolution and structure \cite{2015JATIS...1a4003R}.}

{ \textls[-35]{In this work, we collected the TML based on the light curves from TESS (from MAST Portal\endnote{{\url{https://mast.stsci.edu/portal/Mashup/Clients/Mast/Portal.html}, accessed on 1 May 2024}}, which were processed by the TESS Science Processing Operations Center (SPOC) \cite{SPOC2016,2020RNAAS...4..201C}), AAVSO\endnote{{\url{https://www.aavso.org/}, accessed on 1 June 2024}}, BAV\endnote{{\url{https://www.bav-astro.eu/index.php}, accessed on 1 July 2024}} projects, and the historical literature. The overview of the data from TESS, AAVSO, and BAV is listed in Table \ref{tab:overview}.}  

}

\begin{table}[H]
  \caption{{Overview} of the data from TESS, AAVSO, and BAV.}
  \label{tab:overview}
  \begin{tabularx}{\textwidth}{lCCC}
   \toprule
    \textbf{ID} & \textbf{TESS Sector} & \textbf{Time Spans of AAVSO} & \textbf{Time Span of BAV} \\ 
\midrule       
XX Cyg & {14 (2019), 41 (2021)} & 2008--2023 & 2001--2022 \\ 
YZ Boo & {24 (2020), 50 (2022)} & 2006--2024 & 2000--2016 \\
GP And & {17 (2019)}         & 2008--2023 & 1998--2014 \\
ZZ Mic & {1 (2018), 27 (2020)} & 2006--2007 & ---         \\
\bottomrule
  \end{tabularx}
\\
\footnotesize{Note: All the TESS data used in this work have an exposure time of 120 s.}
\end{table}

{ For the light curves from TESS, we downloaded the Sectors with an exposure time of 120 s and normalized them using the {python package {\tt Lightkurve v2.4.2}}
 \cite{2018ascl.soft12013L}. Then, we used { the module {\tt optimize.curve\_fit} in the {package {\tt SciPy v1.14.0}} \cite{2020SciPy-NMeth}} to fit the data points around each of the peaks in the light curve with a polynomial to determine the time of maximum light. Due to the profiles of the light curves, we used a third- or fourth-order polynomial, generally, and sometimes a fifth-order polynomial, depending on the quality of the data and the goodness of fit. }
 
\textls[-15]{Because Julian Days (JD) are used in the AAVSO data, we converted them into Heliocentric Julian Days (HJD) with the help of an online applet\endnote{{\url{https://doncarona.tamu.edu/apps/jd/}, accessed on 15 June 2024}}.
Moreover, a Monte Carlo simulation was {constructed} to estimate the uncertainties of {each time} of maximum light.}

{For the TML, from the historical literature without uncertainties, we used the following rules to estimate the uncertainties: when the the data value has $n$ decimals, we set the uncertainty as $2\times 10^{-n}$ (i.e., $0.^{\rm d}002$ for $n=3$,  $0.^{\rm d}0002$ for $n=4$).
All the TML in this work {were converted to} BJD with the help of the online applet {\tt hjd2bjd}\endnote{{\url{https://astroutils.astronomy.osu.edu/time/hjd2bjd.html}, accessed on 15 July 2024}}.}

\section{\boldmath{$O-C$}  Analysis}
\label{sec:O-C}

{All the TML used in this work were collected in a data file, which was  uploaded to the Zenodo website (\url{https://zenodo.org/records/14950457}, {accessed on 1 March 2025)}}. 
The indicators of the historical literature in the data file are listed in Appendix \ref{app:01}.

\subsection{$O-C$ Analysis of XX Cyg }
For XX Cyg, a total of 905 TML {were obtained}, {of which} 360  were determined from TESS, 90  were determined from AAVSO, 36  were {determined} from BAV, and 419 of them {were collected} from {the historical literature} of this star. {In order to} calculate the $O-C$ values and their corresponding cycle numbers, we adopted {the linear ephemeris \cite{2012AJ....144...92Y}}                       
\begin{equation}
{\rm HJD_{max}} = 2430671.1102(6) + 0.13486507(1) \times E,
\end{equation}
{to obtain a new linear ephemeris}
\begin{equation}
{\rm BJD_{max}} = 2430671.102(3) + 0.134865132(2) \times E.
\end{equation}

\begin{figure}[H]
   \centering
    \includegraphics[width=0.8\linewidth]{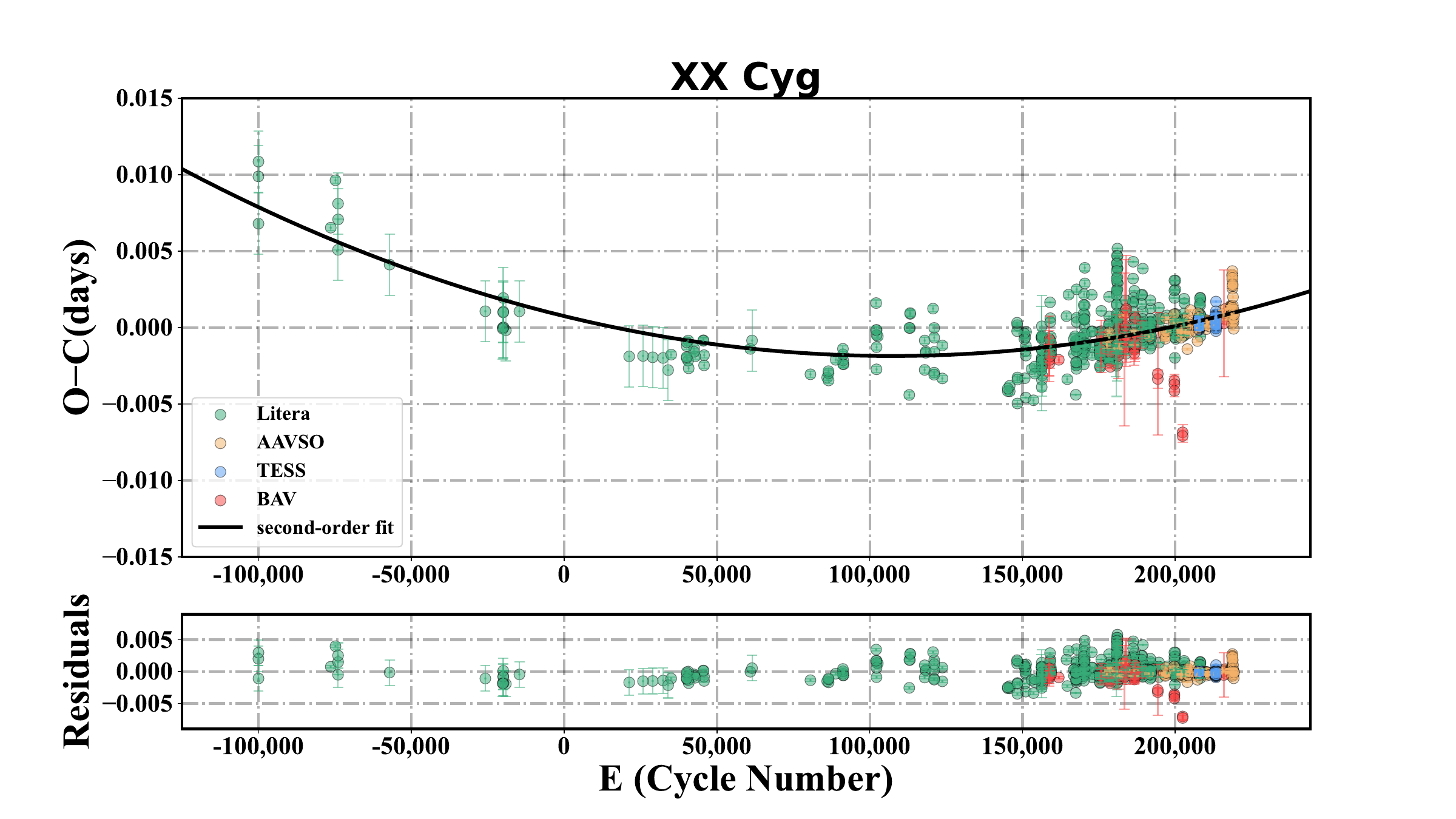}
    \caption{$O-C$ {values} of XX Cyg and the corresponding residuals. The {black} line represents the second-order polynomial fit of the $O-C$ {values}. The different colors of the points represent data from different sources.}

    \label{fig:XXCyg_oc}
\end{figure}

\textls[-25]{{{Then,} 
we fit the $O-C$ values (the residuals of the linear fitting) with a second-order polynomial and obtained the result $O-C = (0.00076\pm 0.00002) +(-4.9\pm0.1)\times10^{-8}\times E + (2.29\pm0.05)\times10^{-13}\times E^{2}$, which has a linear period variation rate of $\dot{P}/P = (9.2 \pm 0.2) \times 10^{-9} \ \mathrm{yr^{-1}}$. The fitting results of the $O-C$ values and corresponding residuals are shown in Figure \ref{fig:XXCyg_oc}.}}

\subsection{$O-C$ Analysis of YZ Boo }
\textls[-15]{For YZ Boo, a total of 736 TML {were obtained, of which} 434 were determined from TESS, 28 were determined from AAVSO, 63  were determined from BAV, and 211{ were collected} from {the historical literature}. {In order to} calculate the $O-C$ values and their corresponding cycle numbers, we adopted {the linear ephemeris \cite{2018RAA....18....2Y}}}
\begin{equation}
{\rm HJD_{max}} = 2442146.3552(2) + 0.104091579(2) \times E,
\end{equation}
{to obtain a new linear ephemeris}
\begin{equation}
{\rm BJD_{max}} = 2446624.3756(1) + 0.1040915838(8) \times E.
\end{equation}

\begin{figure}[H]
   \centering
    \includegraphics[width=0.8\linewidth]{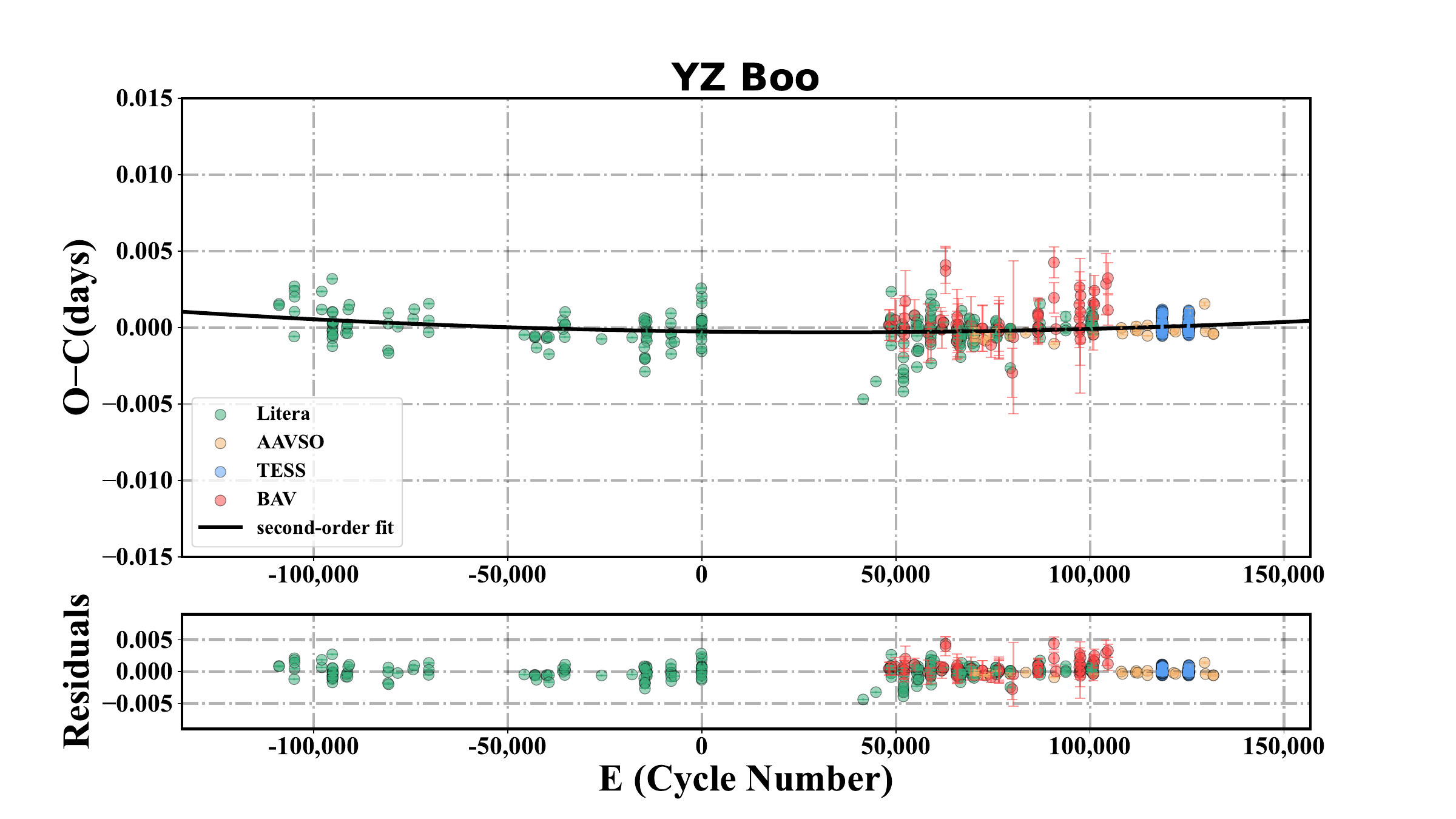}
    \caption{$O-C$ {values} of YZ Boo and the corresponding residuals. The {black} line represents the second-order polynomial fit of the $O-C$ {values}. The different colors of the points represent data from different sources.}
    \label{fig:YZBoo_oc}
\end{figure}

{{Then,} we fit the $O-C$ values (the residuals of the linear fitting) with a second-order polynomial and obtained the result $O-C = (-0.00025\pm 0.00001) +(-3.1\pm0.2)\times10^{-9}\times E + (4.8\pm0.3)\times10^{-14}\times E^{2}$.}
{The obtained linear period variation rate is $\dot{P}/P= (3.2\pm 0.2)\times 10^{-9}\ \mathrm{yr}^{-1}$, which is similar to the result in Ref. \cite{2018RAA....18....2Y}. The fitting results of the $O-C$ values and corresponding residuals are shown in Figure \ref{fig:YZBoo_oc}.}

\subsection{$O-C$ Analysis of GP And}
For GP And, a total of 591 TML {were obtained (from 1970 to 2023), of which} 268 of  were determined from TESS, 33 were determined from AAVSO, 92 were determined from BAV, and 198 of them {were collected} from {the historical literature}. {In order to} calculate the $O-C$ values and their corresponding cycle numbers, we adopted {the linear ephemeris \cite{2011AJ....142..100Z}}
\begin{equation}
    {\rm HJD_{max}} = 2441909.4917(7) + 0.07868269(8) \times E,
\end{equation}
{to obtain a new linear ephemeris}
\begin{equation}
    {\rm BJD_{max}} = 2450360.4942(1) + 0.0786828064(9)\times E.
\end{equation}

{{Then,} we fit the $O-C$ values (the residuals of the linear fitting) with a second-order polynomial and obtained the result $O-C = (-0.00214\pm 0.00002) +(-1.106\pm0.008)\times10^{-8}\times E + (3.57\pm0.03)\times10^{-13}\times E^{2}$.}

\begin{figure}[H]
    \centering
    \includegraphics[width=0.8\linewidth]{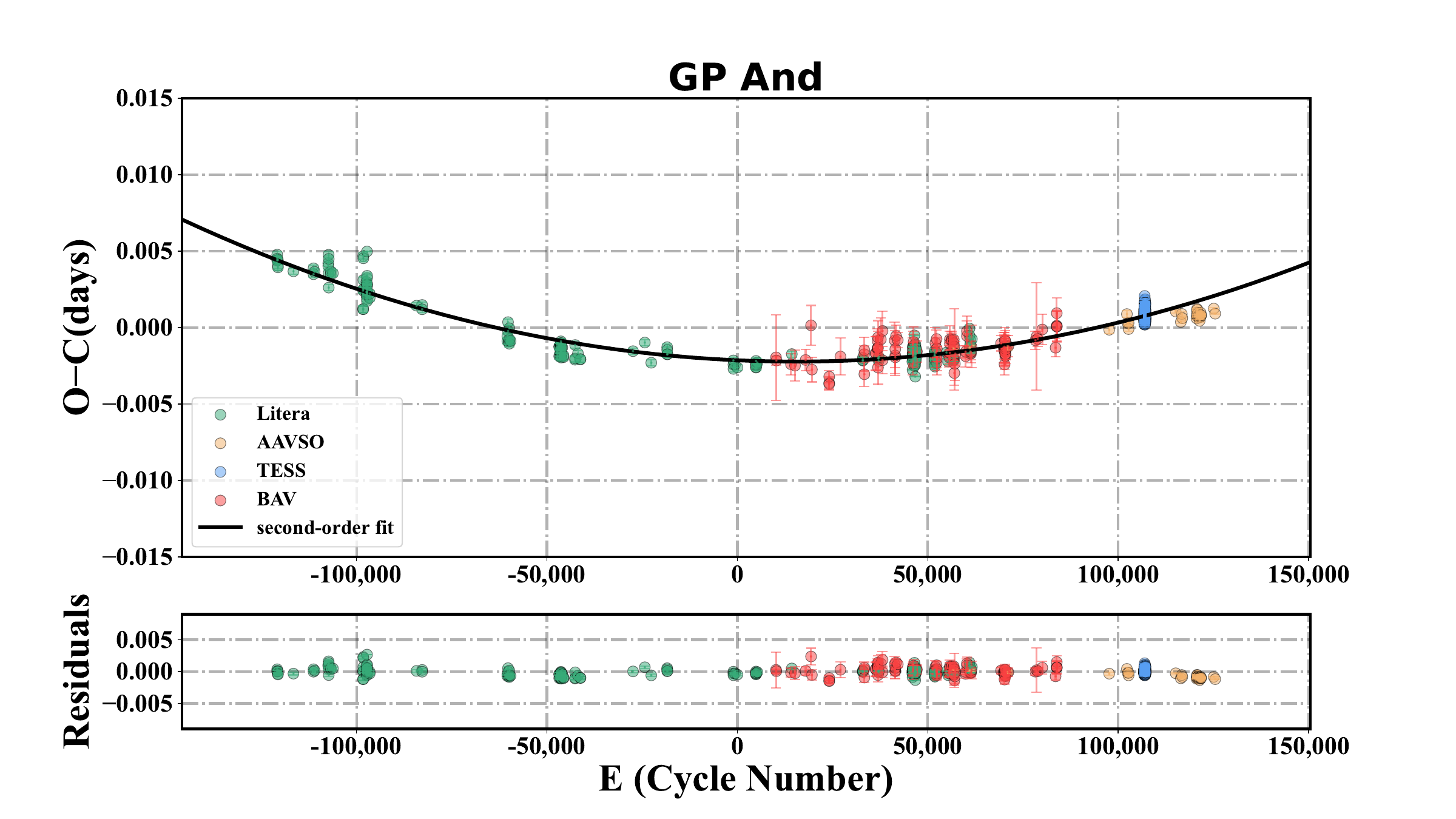}
    \caption{$O-C$ {values} of GP And and the corresponding residuals. The {black} line represents the second-order polynomial fit of the $O-C$ {values}. The different colors of the points represent data from different sources.} 
    \label{fig:GPAnd_oc}
\end{figure}

{The obtained linear period variation rate is $\dot{P}/P= (4.22\pm 0.03) \times 10^{-8}\ \mathrm{yr}^{-1}$, and the fitting results of the $O-C$ values and corresponding residuals are shown in Figure \ref{fig:GPAnd_oc}.}

\subsection{$O-C$ Analysis of ZZ Mic}
{For ZZ Mic}, we determined 667 TML from TESS and 2 from AAVSO. Moreover, we also collected 48 TML from {the historical literature}. 
{Based on the above 717 TML, we determined the new linear ephemeris as}
\begin{equation}
{\rm BJD_{max}} =  2449996.6664(5) +  0.0671791806(8)\times E. 
\end{equation}

{{Then,} we fit the $O-C$ values (the residuals of the linear fitting) with a second-order polynomial and obtained the result $O-C = (0.00205\pm 0.00002) +(2.37\pm0.02)\times10^{-10}\times E + (-1.27\pm0.01)\times10^{-13}\times E^{2}$.}

\begin{figure}[H]
    \centering
    \includegraphics[width=0.8\linewidth]{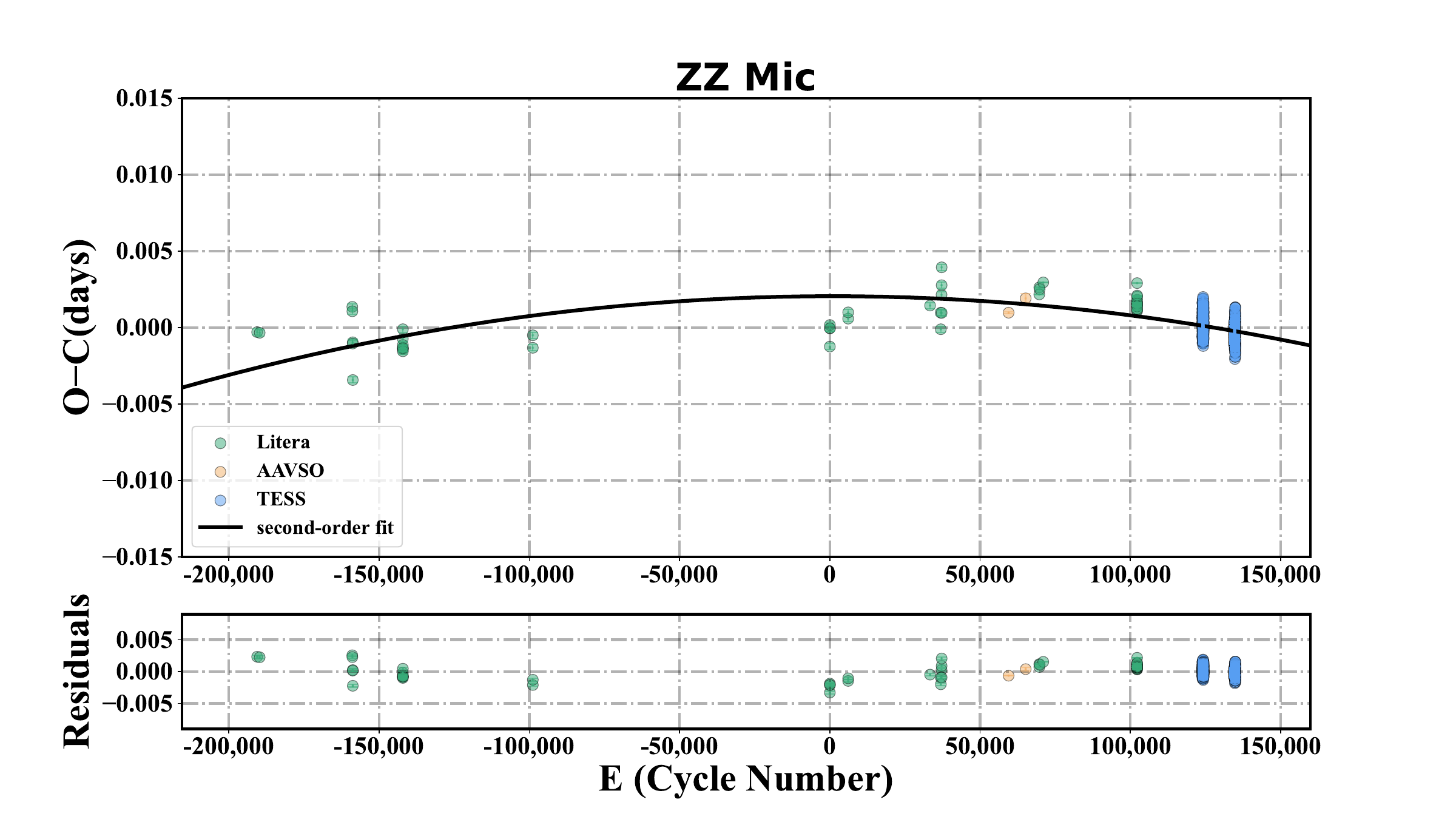}
    \caption{$O-C$ {values} of ZZ Mic and the corresponding residuals. The {black} line represents the second-order polynomial fit of the $O-C$ {values}. The different colors of the points represent data from different sources.}
    \label{fig:ZZMic_oc}
\end{figure}

{The obtained linear period variation rate is $\dot{P}/P= (-2.06\pm 0.02) \times 10^{-8}\ \mathrm{yr}^{-1}$, and the fitting results of the $O-C$ values and corresponding residuals are shown in Figure \ref{fig:ZZMic_oc}.}

\section{Discussion}
\label{sec:discussion}

\subsection{Discussion of XX Cyg}
\textls[-15]{The {linear} period variation rate we {obtained} in this work was $\dot{P}/P = (9.2 \pm 0.2) \times 10^{-9}\ \mathrm{yr}^{-1}$, {which is similar to the theoretical evolutionary results in ref. \cite{2012AJ....144...92Y}. In this post-mean sequence (post-MS) evolutionary stage,} the envelope {absorbs} energy produced by the {hydrogen burning shell} and then {expands}, {which leads to an expansion of the star.}  
As it well known, the period of the fundamental mode in HADS mainly {depends} on the mean density of the star itself. So the increase in the {stellar} radius means an increase in the period. }

In some earlier research (see, e.g., ref. \cite{2000IBVS.4950....1K}), {researchers} claimed that XX Cyg experienced a sudden period variation in 1942. However, even though the jumping of period variation is allowed by stellar evolution theory \cite{1999NewAR..43..441B}, {our result indicates that it should be due to the limited data points}.   

In addition, some {other} HADS that were suspected to exhibit period jumps have now been proved to have a {continuous} period variation, or the {previously considered `jump' was actually caused by the light-travel time effect (LTTE) (such as CY Aqr \cite{1995PASP..107..225P,Fang_2016})}.

\subsection{Discussion of YZ Boo}
Different from {the other} three stars in this work, we found that the second-order polynomial trend in the $O-C$ diagram of YZ Boo was not very significant. Therefore, an F-test was performed {to test the advantage of the second-order fitting comparing to the linear one. Based on the routine from Statology\endnote{{\url{https://www.statology.org/f-test-python/}, accessed on 15 September 2024}}, we used the residuals of the linear and second-order fittings to calculate the $p$ value. It produced $p \approx 0.26$, and then we obtained $1-p \approx 0.74$, which indicates that the goodness-of-fit of the second-order fitting was not significant.} As a result, we {cannot} conclude that YZ Boo exhibited {significant} period variation in 1955--2024. This result is {similar to} the conclusion in ref. \cite{2018RAA....18....2Y}.

\textls[-15]{{However, if we take the linear period variation rate of YZ Boo obtained in this work seriously, it is consistent with the theoretical calculations of the period variation rate in ref. \cite{2018RAA....18....2Y}, which indicates that YZ Boo is in the post-MS evolutionary stage. In the case of YZ Boo, we need more data to obtain results with  higher confidence.}}

\subsection{Discussion of GP And}
{Although the linear period variation rate of GP And ($\dot{P}/P = (4.22\pm 0.03) \times 10^{-8}\ \mathrm{yr}^{-1}$) falls into the range of the theoretical prediction ($10^{-10}\ \mathrm{yr^{-1}}$ to $10^{-7}\ \mathrm{yr^{-1}}$) and indicates the star is in the post-MS evolutionary stage, a detailed stellar evolutionary model for this star is also needed to explain the exact value of $\dot{P}/P$ and its evolutionary stage.}

\textls[-20]{{ Ref. \cite{1993DSSN....6...10R} calculated the linear period variation rates of GP And (including both MS and post-MS models) in three cases, and all the results were at the order of $10^{-9}\ \mathrm{yr}^{-1}$. Ref. \cite{pop2003period} estimated a theoretical value of about $2\times10^{-9}\ \mathrm{yr}^{-1}$ from the results in ref.~\cite{1998A&A...332..958B}.}
{On the other hand, the observed linear period variation rate of GP And was $\dot{P}/P = (5.49\pm 0.1) \times 10^{-8}\ \mathrm{yr}^{-1}$~\cite{2011AJ....142..100Z} and $ 5.39 \times 10^{-8}\ \mathrm{yr}$~\cite{2011Ap&SS.333..125B} in some recent literature. }}

{Why are the theoretical values of the period variation rate of GP And we mentioned above much larger than the observed ones?} Here we give two {possibilities}: (i) First, { more accurate and detailed theoretical evolutionary models need to be constructed for  GP And, which could give us a more reasonable theoretical prediction of $\dot{P}/P$}; (ii) {Second, an accumulation of the TML in a longer time-scale is also needed to test whether the observed $\dot{P}/P$ is caused by LTTE.} Because a {quasi-sinusoidal} curve is locally similar to a polynomial curve, the observed {second-order polynomial trend could be a superposition of a quasi-sinusoidal trend (from LTTE) and a weak second-order polynomial trend (from stellar evolution) (see, e.g., refs. \cite{2018MNRAS.473..398L,2020ApJ...904....5X}). }

\subsection{Discussion of ZZ Mic}
The situation in ZZ Mic is more complex than that of the other three stars. Historically, some researchers obtained an increasing period \cite{2009PASP..121..478K} and even employed a third-order polynomial to fit the $O-C$ values. 
{The negative linear period variation rate obtained in this work indicates that ZZ Mic has a decreasing period. The different results might come from the lack of TML in cycles from $-100,000$ to $0$, which prevents us from obtaining high-confidence results on the period variation of ZZ Mic.}

{If we ascribe the decreasing period to stellar evolution, it demands that ZZ Mic is in an overall contraction evolutionary stage.}
In this stage, the star has {exhausted} all the hydrogen in {its} core and starts to contract due to gravity. 
{Although some works reported pre-MS stars (which show delta Scuti-type pulsations) can also have negative period variation \cite{1972ApJ...171..539B,2014A&A...568A..32D,doi:10.1126/science.1253645}, the linear period variation rate of ZZ Mic is so small that it can only be a post-MS star \cite{1998A&A...332..958B}.}

\begin{figure}[H]
    \centering
    \includegraphics[width=0.8\linewidth]{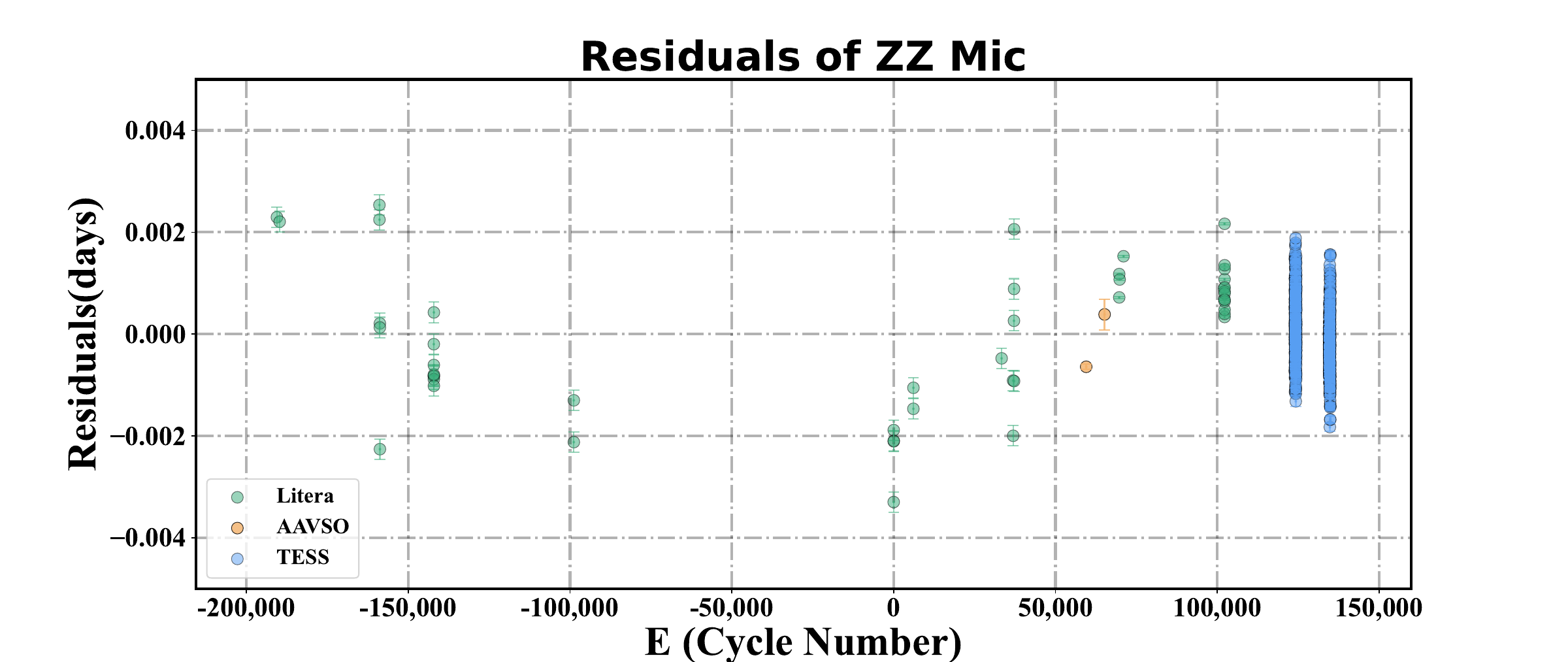}
    \caption{{Residuals} of the second-order fitting of the $O-C$ values from ZZ Mic.}     \label{fig:ZZMic_res}
\end{figure}

{ The fitting residuals of the O-C values of ZZ Mic are presented in Figure \ref{fig:ZZMic_res}, in which we can find a potential periodic trend. However, the data points were not sufficient to cover a complete cycle. 
As the  TML continues to be accumulated from different observations, we expect to solve out the potential orbital parameters via the $O-C$ values affected by LTTE in this multi-star system (see, e.g., refs. \cite{2020ApJ...904....5X,2019PASP..131f4202Z}) and study some other interesting mechanisms like the magnetic braking effect (which is an important physical mechanism that influences material accretion, spin-down, and evolutionary paths of binary stars \cite{Deng2020,Deng2021}) in other types of multi-star systems.} 
\vspace{-6pt}

\section{Conclusions}
\label{sec:conclusions}
In this {work}, we collected TML of four HADS over several decades {and obtained the linear period variation rates of these stars via the $O-C$ analysis as follows:} (i) XX Cyg, $\dot{P}/P = (9.2 \pm 0.2) \times 10^{-9}\ \mathrm{yr}^{-1}$; (ii) YZ Boo, $\dot{P}/P = (3.2\pm 0.2)\times 10^{-9}\ \mathrm{yr}^{-1}$; (iii) GP And, $\dot{P}/P = (4.22\pm 0.03) \times 10^{-8}\ \mathrm{yr}^{-1}$; (iv) ZZ Mic, $\dot{P}/P = (-2.06\pm 0.02) \times 10^{-8}\ \mathrm{yr}^{-1}$.

For XX Cyg and YZ Boo, the period variation rates we {obtained} are similar to the results in the {latest} research, which could be ascribed to stellar evolution. 
{For GP And, although the observed period variation rate is in the range of the general theoretical prediction from stellar evolution, some discrepancy appears between the the observed value and some previous specialized studies on this star. More in-depth studies are necessary in the future, based on more observations.}

For ZZ Mic, {we obtained a negative linear period variation rate}, which is {different from previous studies and indicates that the star is in an overall contraction evolutionary stage}. 
{Moreover, we found a potential periodic (quasi-sinusoidal) trend in the residuals of $O-C$ diagram, which might be caused by LTTE in a multi-star system.}

{More observations in the future will provide us further opportunity to study these HADS in-depth, including their evolutionary stages and their nature in multi-star systems. }

\vspace{+6pt}

\authorcontributions{{Conceptualization, J.S.N and H.F.X; methodology, J.S.N and H.F.X; software, T.F.M; validation, T.F.M, J.S.N, and H.F.X; formal analysis, T.F.M; investigation, T.F.M; resources, J.S.N; data curation, T.F.M.; writing---original draft preparation, T.F.M.; writing---review and editing, J.S.N and H.F.X; visualization, T.F.M; supervision, J.S.N and H.F.X; project administration, J.S.N and H.F.X; funding acquisition, H.F.X. All authors have read and agreed to the published version of the manuscript.}} 

\funding{This research was supported by the National Natural Science Foundation of China (NSFC: No. 12303036).}

\dataavailability{All the data obtained in this work can be found on the Zenodo website (\url{https://zenodo.org/records/14950457}, {accessed on 1 March 2025).}}   

\acknowledgments{H.F.X. acknowledges support from the National Natural Science Foundation of China (NSFC) (No. 12303036). 
All the authors acknowledge the TESS Science team and everyone who has contributed to making the TESS mission possible. We also acknowledge with thanks the variable star observations from the AAVSO International Database contributed by observers worldwide and used in this research.}
\conflictsofinterest{The authors declare no conflicts of interest.} 

\appendixtitles{yes} 
\appendixstart
\appendix
\section[\appendixname~\thesection. Times of Maximum Light Used in This Work]{Times of Maximum Light Used in This Work}
\label{app:01}
All the TML used in this work were collected in a file and uploaded on the Zenodo website (doi: 10.5281/zenodo.14950457, \url{https://zenodo.org/records/14950457}).

The indicators of the historical literature in the data file are listed as follows.

For XX Cyg: 
(1) \cite{1906ApJ....23...79P}; (2)\cite{1915ApJ....42..148S}; (3) \cite{1929PAllO...7....1J}; (4) \cite{1938AN....267..137K}; (5) \cite{1936AN....258..329D}; (6) \cite{1981CoKon..75....1S}; (7) \cite{1981CoKon..75....1S}, from unpublished data of L. Detre; (8) \cite{1966CoLPL...5...71F}; (9) \cite{1982IBVS.2161....1R}; (10) \cite{1980PASP...92..195M}; (11) \cite{Joner_1982}; (12) \cite{1986PASP...98..504S}; (13) \cite{kim1994vr}; (14) \cite{rodriguez1993simultaneous}; (15) \cite{1996vsr..conf...18A}; (16) \cite{2003PASP..115..212B}; (17) \cite{2000IBVS.4912....1A}; (18) \cite{zhou2002amplitude}; (19) \cite{2000IBVS.4950....1K}; (20) \cite{2005IBVS.5643....1H}; \cite{2005IBVS.5657....1H}; (21) \cite{2006IBVS.5701....1K}; (22) \cite{2006IBVS.5731....1H}; (23) \cite{2006IBVS.5684....1B}; (24) \cite{2006IBVS.5731....1H}; (25) \cite{2009MNRAS.394..995D}; (26) \cite{2009OEJV..114....1F}; (27) \cite{2011PASP..123...26C}; (28) \cite{2012AJ....144...92Y}; (29) \cite{2021JRASC.115..238W}.

For YZ Boo:
(1) \cite{1955PASP...67..354E}; (2) \cite{1957MmSAI..28...13B}; (3) \cite{1959ApJ...130..539S}; (4) \cite{1961MmSAI..32....7B}; (5) \cite{1964ApJ...140..694H}; (6) \cite{1966CoLPL...5...71F}; (7) \cite{gieren1974three}; (8) \cite{1976PhDT.........1L}; (9) collected from \cite{2008JRASC.102..134W}, the original source cannot be found; (10) \cite{1981CoKon..75....1S}; (11) \cite{1983PASP...95..433J}; (12) \cite{1985AcASn..26..297J}; (13)~\cite{1985PASP...97.1172P}; (14) \cite{1986IBVS.2963....1H}; (15) \cite{kim1994vr}; (16) \cite{1999IBVS.4712....1A}; (17) \cite{2000IBVS.4912....1A}; (18) \cite{derekas2003photometric}; (19) \cite{2003IBVS.5485....1A}; (20) \cite{jin2003multiband}; (21)  \cite{2005IBVS.5643....1H}; (22) \cite{zhou2006stability}; (23)~\cite{2006IBVS.5701....1K}; (24) \cite{2006IBVS.5731....1H}; (25) \cite{2008JRASC.102..134W}; (26) \cite{2018RAA....18....2Y}.

For GP And:
(1) \cite{1976MitVS...7..137S}; (2) \cite{gieseking1979photoelectric}, the TML were derived from \cite{rodriguez1993simultaneous}; (3) \cite{1978IBVS.1517....1E}, the TML were derived by \cite{rodriguez1993simultaneous}; (4) \cite{rodriguez1993simultaneous}; (5) \cite{burchi1993photoelectric}; (6) \cite{1995AJ....109.1239S}; (7) collected from \cite{2006IBVS.5718....1S}; (8) \cite{2006IBVS.5718....1S}; (9) \cite{2011AJ....142..100Z}.

For ZZ Mic:
(1) \cite{1961Obs....81...25C}; (2) \cite{1968AJ.....73..500L}; (3) \cite{1971ApJ...165..365C}; (4) \cite{1978MNRAS.184...11B}; (5) \cite{2009PASP..121..478K}; (6) \cite{2015JAVSO..43...50A}.

\begin{adjustwidth}{-\extralength}{0cm}
\printendnotes[custom]

\reftitle{References}

\isAPAandChicago{}{%

}{}

\PublishersNote{}

\end{adjustwidth}

\end{document}